\documentclass[twocolumn,preprintnumbers,amsmath,amssymb]{revtex4}

\usepackage{graphicx}  
\begin{document}

\title{
NEGATIVE  TERAHERTZ  CONDUCTIVITY IN REMOTELY  DOPED  GRAPHENE 
BILAYER HETEROSTRUCTURES
}
\author{V. Ryzhii,$^{1,2}$ M. Ryzhii,$^3$  V. Mitin,$^{4}$ M. S. Shur,$^5$
and T. Otsuji$^1$
 }
\affiliation{
$^1$ Research Institute of Electrical Communication, Tohoku University,  Sendai 980-8577, Japan\\ 
$^2$ Institute of Ultra High Frequency Semiconductor Electronics of RAS,
and  Center for Photonics and Infrared Engineering, 
Bauman Moscow State Technical University, Moscow 111005, Russia\\ 
$^3$ Department of Computer Science and Engineering, University of Aizu,
Aizu-Wakamatsu 965-8580, Japan\\
$^4$ Department of Electrical Engineering, University at Buffalo, SUNY, Buffalo, New York 1460-1920, USA\\
$^5$ Departments of Electrical, Electronics, and Systems Engineering and Physics, Applied Physics, and Astronomy, Rensselaer Polytechnic Institute, Troy, NY 12180, USA
}
\begin{abstract}
Injection or optical generation  of electrons and holes in  graphene bilayers (GBLs) can
result in the interband population inversion enabling the terahertz (THz) radiation lasing. 
The intraband radiative processes compete with the interband transitions. We demonstrate that remote doping enhances the indirect interband generation of photons in the proposed GBL heterostructures. Therefore such remote doping helps surpassing the intraband (Drude) absorption and results in large absolute values of the negative dynamic THz conductivity in a wide range of frequencies at elevated (including room) temperatures. The remotely doped GBL heterostructure THz lasers are expected to achieve higher THz gain compared to previously proposed GBL-based THz lasers.
\end{abstract}
\maketitle


\section{Introduction}
 
 The population inversion created by injection or optical pumping in graphene layers (GLs) and graphene bilayers (GBLs)~\cite{1} results in the negative dynamic conductivity in the terahertz (THz) range of frequencies~\cite{2,3} and enables the graphene-based THz lasers~\cite{2,3,4,5}.
 Different research groups have  demonstrated the THz gain in the pumped GLs ~\cite{6,7,8,9,10,11,12,13,14,15,16,17} (see, also
Ref.~\cite{17}).
GLs and GBLs 
  can serve as the active regions of the 
THz lasers with the Fabry-Perot resonators, dielectric waveguides,  slot-lines including 
 the plasmonic lasers~\cite{8,19,20,21,22,23,24}. 
 The GL and GBL heterostructure lasers can 
  operate in a wide frequency range, including the
 6 to 10 THz range, where using materials like A$_3$B$_5$ is hindered by the optical phonon effects~\cite{25,26,27}.
The THz lasing in GL and GBL structures  is possible if the contribution of the interband radiative transitions  to the  real part of the dynamic conductivity ${\rm Re}~\sigma$ in a certain range of frequencies surpasses the contribution of the intraband radiative processes associated with  the Drude
 losses. 
The interband  
 transitions include the direct (with the conservation of the electron momenta)
and the indirect transitions (accompanied with the variation of  the electron momentum due to  scattering). 
The limitations imposed by the momentum and energy conservation laws allow only for the indirect intraband radiative transitions.
In the GLs and GBLs with sufficiently long carrier momentum relaxation time $\tau$,
the direct interband radiative transitions dominate over the indirect intraband transitions and 
${\rm Re}~\sigma < 0.$
Recently~\cite{28,29}, we demonstrated that in GLs and GBLs with primarily long-range disorder scattering,  the indirect interband transitions can 
prevail
over the indirect intraband transitions leading to 
fairly large absolute values of the negative conductivity~$|\rm~Re\sigma|$.
These values could 
 exceed the fundamental limits for the direct transitions, which are  $\sigma_Q = e^2/4\hbar$
for GLs and $2\sigma_Q$ for GBLs~\cite{1} (here $e$ is the electron charge and $\hbar$ is the Planck constant). 
As a result, in the GL and GBL structures with a long-range disorder one can expect that
the condition  ${\rm Re}~\sigma < 0$ could be  fulfilled in a wider range of the THz frequencies,
including   the low boundary of the THz range.
When the momentum relaxation is associated with long-range scattering mechanism,
moderate values of $\tau$ might be sufficient for achieving ${\rm~Re}~\sigma < 0$, especially in GBLs where the 
density of states (DoS) $\rho(\varepsilon)$ as a function of  energy $\varepsilon$ near the band edges ($\rho(\varepsilon)\simeq const$) in the latter is considerably  larger than in GLs
(where $\rho(\varepsilon) \propto |\varepsilon|$).

This paper deals with 
 the GBL heterostructures with a  remote impurity layer (RIL)
 located at some distance from the GBL plane and incorporating donors and acceptors or donor and acceptor clusters. We consider the GBL-RIL heterostructure
 as an active region of an  injection THz laser.
We show that  the electron and hole injection from the side n- and p-contact regions into the GBL-RIL heterostructures leads to
the population inversion enabling the enhanced negative dynamic conductivity, which, in turn, can result in lasing in the THz range.   
 Figure~1 shows the schematic view of an injection laser based on  the GBL-RIL heterostructure at the reverse bias voltage $V$.   
The remote (selective) doping can markedly affect the scattering of electrons and holes (see, for example~\cite{30,31}) and, hence,
the  indirect radiative transitions.   
The GL and GBL heterostructures using remote doping were already fabricated and used for  device applications~\cite{32,33}.
The heterostructures  with the selectively doped GBL and RIL demonstrate the following   distinct features. First, the net  electron and hole densities  determined by both the pumping and RIL and the electron and hole quasi-Fermi energies, $\varepsilon_{F,e}$ and  $\varepsilon_{F,h}$, are generally not equal.
Second, 
 relatively large  scale  of the donor and acceptor density fluctuations   result    in smooth (long-range) spatial
variations of the potential in the GBL  created by  the RIL, separated by distance $d$
from the GBL. 
 This could  result in the electron and hole scattering with a  relatively small momentum change and,
 consequently, increase in the contribution of the indirect interband transitions in comparison with the interaband transitions.
Thus, the main role of the RIL is not to induce extra carrier in the GBL (as in high-electron mobility transistors) but to   provide  the specific mechanism
of the electron and hole scattering reinforcing the indirect interband radiative transitions. 
We calculate  the spectral characteristics of real part of the dynamic conductivity ${\rm Re}~\sigma$ in the pumped GBL-RIL heterostructures  as a functions of the doping level and the spacer thickness $d$ and compare the obtained characteristics with those of the undoped heterostructures.
This analysis reveals the conditions  for the selective doping
enabling   the negative dynamic conductivity in a wide
 frequency range, including the frequencies of a few THz.
This might open new prospects 
of GBL heterostructures for the efficient  THz lasers based on the GBL heterostructures.
 
 \section{Device model}

\begin{figure}[t]
\centering
\includegraphics[width=6.0cm]{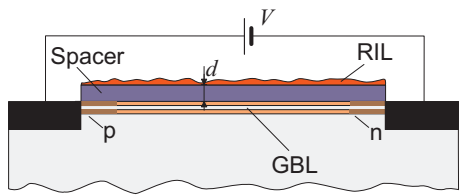}
\caption{Schematic view  of an injection  laser  based on GBL-RIL heterostructure with  the GBL doped with acceptors and
the RIL doped with both donors and acceptors and with
the chemically  doped n- and p- injectors.  
}
\end{figure}

\begin{figure}[t]
\centering
\includegraphics[width=6.0cm]{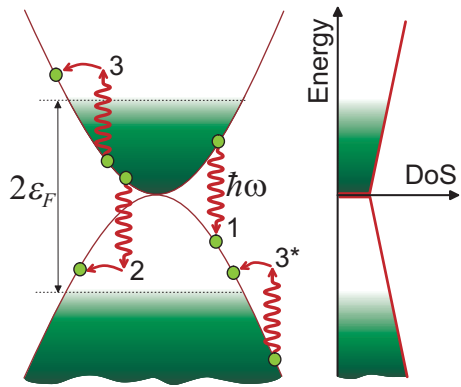}
\caption{Energy band diagram  of a pumped  GBL  (left panel) and energy dependency of its DoS (right panel). Arrows correspond to direct and indirect interband  transitions with the photon emission  (wavy arrow "1" and wavy + smooth arrows "2")  as well as to indirect   intraband transitions with the photon absorption in the conduction and valence bands (wavy + smooth arrows "3" and "3*", respectively). 
}
\end{figure}

We study the  GBL-RIL heterostructures pumped  via the injection from the side
p- and n- contacts shown in  Fig.~1. 
In the very clean pumped GBLs (or GLs)  with a fairly high carrier mobility  the intraband (Drude absorption) at the radiation frequencies above one THz should not play a dominant role.
Hence, adding the selective doping might not be needed.
However, 
 in the GBLs with a relatively low carrier mobility caused by unavoidable (residual) charge impurities
 and imperfections, the  RIL induced scattering could  result in a substantial compensation of the Drude absorption
and, therefore, enhancement of  the negative dynamic conductivity 
In the following, for definiteness, we consider the  GBL-RIL heterostructures  with
 the GBL 
  doped with acceptors (with the density $\Sigma_A$) and the RIL formed by partially compensated  donors (with the density $\Sigma_{D,R}$) and
acceptors (with the density $\Sigma_{A,R}$). Hence  $\Sigma_{D,R} - \Sigma_{A,R} \simeq \Sigma_A$.
We further assume that in these heterostructures, the RIL might comprise the clusterized acceptors and donors (with the correlated charge defects forming the charged clusters with the charge $Z_ce > e$ 
and the characteristic size $l_c$, where $e$ is the electron charge).
In the latter case, we assume that  the cluster  density is equal to 
 $(\Sigma_{A,R} + \Sigma_{D,R})/Z_c $.
Such  GBL-RIL heterostructures  are quasi-neutral
 even  in the absence of the pumping. 
Pumping  
 results in the formation in the GBL of the two-dimensional (2D) electron and hole gases with the equal densities $\Sigma_e =\Sigma_h\ = \Sigma$ and the quasi-Fermi energies $\varepsilon_{F,e} = \varepsilon_{F,h}= \varepsilon_{F}$. 
Figure~2 shows the GBL band diagram under pumping.
In Fig.~2,  the arrows indicate different radiative transitions: vertical interband transitions,
indirect intraband transitions accompanied by the electron scattering on the impurities, phonons,
and holes (leading to the Drude absorption of radiation), and the interband indirect transitions.

The transverse electric field arising in the spacer due to the features of the doping can lead to the local opening of the 
energy gap~\cite{1,34}. However, as shown below, for the doping levels under consideration, the energy gap 
expected to be small, and  
the band opening and 
and the energy spectrum nonparabolicity  
will be disregarded~\cite{34,35,36}. The energy spectrum is assumed to be
 $\varepsilon_{\bf p} = \pm p^2v_W^2/\gamma_1 = \pm p^2/2m$, ~\cite{1}, where  $\gamma_1 \simeq 0.35 - 0.43$~eV
 is the inter-GL overlap integral in GBLs, $v_W \simeq 10^8$~cm/s is the characteristic velocity, and $m = \gamma_1/2v_W^2$ is the effective mass of electrons and holes.
%

\section{Main equations}

The real part of the net dynamic (THz) conductivity in the in-plane direction, ${\rm Re}~\sigma$, of GLs and GBLs    comprises the contributions 
from the direct (vertical),  ${\rm Re}~\sigma_d$, and two types of the indirect, 
${\rm Re}~\sigma^{inter}_{ind}$ and  ${\rm Re}~\sigma^{intra}_{ind}$, transitions, respectively:

\begin{equation}\label{eq1}
 {\rm Re}~\sigma  = {\rm Re}~(\sigma_d + \sigma_{ind}^{inter}  + \sigma_{ind}^{intra}).
\end{equation}
The first term in Eq.~(1) corresponds to the transitions of the type "1" in Fig.2,
the second term corresponds to the type "2", and the third term - to the types "3" and "3*" transitions, respectively. The last term in the right-hand side of Eq.~(1) corresponds  to the processes responsible  the Drude absorption.

Due to high frequencies of the  inter-carrier scattering under sufficiently strong pumping,
the energy distributions of the pumped carriers
are characterized by the Fermi distribution functions $f_v(\varepsilon_{\bf p})$ and 
$f_c(\varepsilon_{\bf p})$, where $\varepsilon_{\bf p}$ is the dispersion relationship for the 2D carriers  in GBLs, 
 with the quasi-Fermi energy, $\varepsilon_{F} \simeq \pi\hbar^2\Sigma/2m$   and the effective temperature $T$. 

The calculations of  the GBL dynamic conductivity in the in-plain direction associated with the direct inter-band transitions accounting for the GBL energy spectra, following the approach developed in ~\cite{37,38} (see, also~\cite{5}),
yield:
\begin{multline}\label{eq2}
 {\rm Re}~\frac{\sigma_d}{\sigma_Q} = \frac{(\hbar\omega + 2\gamma_1)}{(\hbar\omega +\gamma_1)}
 \displaystyle\tanh\biggl(\frac{\hbar\omega - 2\varepsilon_F}{4T} \biggr)
 \\
 \simeq 2\displaystyle\tanh\biggl(\frac{\hbar\omega - 2\varepsilon_F}{4T} \biggr).
\end{multline}

Applying the Fermi golden rule for the indirect intra- and
interband electron radiative transitions, we obtain at the following formulas
 for the respective components of the dynamic conductivity:
 %
\begin{multline}\label{eq3}
{\rm Re}~\frac{\sigma_{ind}^{intra}}{\sigma_Q}\\ =  \frac{4\pi\,g}{\hbar\omega^3}
\sum_{{\bf p}, {\bf p}^{\prime}} 
|V({\bf p} - {\bf p}^{\prime})|^2 u^{\lambda\lambda}_{{\bf p}{\bf p}^{\prime}}({\bf v}_{{\bf p}^{\prime}} -
{\bf v}_{{\bf p}})^2\delta(\hbar\omega + \varepsilon_{{\bf p}} - \varepsilon_{{\bf p}^{\prime}})\\
\times\biggl\{\frac{1}{1 + \exp [(\varepsilon_{\bf p} - \varepsilon_F)/T]}
- \frac{1}{1 + \exp [(\varepsilon_{\bf p^{\prime}} - \varepsilon_F)]}
\biggr\}
,
\end{multline}

\begin{multline}\label{eq4}
{\rm Re}~\frac{\sigma_{ind}^{inter}}{\sigma_Q}\\ =  \frac{2\pi\,g}{\hbar\omega^3}
\sum_{{\bf p}, {\bf p}^{\prime}}
|V({\bf p} - {\bf p}^{\prime})|^2 
u^{cv}_{{\bf p}{\bf p}^{\prime}}({\bf v}_{{\bf p}^{\prime}} +
{\bf v}_{{\bf p}})^2\delta(\hbar\omega - \varepsilon_{{\bf p}} - \varepsilon_{{\bf p}^{\prime}})\\
\times\biggl\{\frac{1}{1 + \exp [(\varepsilon_{\bf p} - \varepsilon_F)/T]}
- 
\frac{1}{1 + \exp [(\varepsilon_{\bf p^{\prime}} + \varepsilon_F)/T]}
\biggr\}. 
\end{multline}
%
%
Here  $g = 4$ is the spin-valley degeneracy factor, $\varepsilon_{\bf p} $,
${\bf v}_{{\bf p}} = 2{\bf p}v_W^2/\gamma_1 = {\bf p}/m$
is the  velocity of electrons and holes with the momentum ${\bf p}$ in  GBLs, 
$V({\bf q})$ is the ${\bf q}$-the Fourier component of the scattering potential, $u^{\lambda\lambda^{\prime}}_{{\bf p}{\bf p}^{\prime}} = (1 + \lambda\lambda^{\prime}\cos 2\theta_{{\bf p}{\bf p}^{\prime}})/2$ is the overlap between the envelope wave functions in GBLs, and $\lambda = \pm 1$ is the index indicating the conduction and valence bands. The extra factor of 2 in Eq.~(3) accounts for
the intraband electron and hole contributions.

Equation~(3) yields the well-known result for the dynamic Drude conductivity in the high-frequency limit $\Gamma = \hbar/\tau < \hbar\omega < \varepsilon_F$, 
where $\Gamma$ is the broadening of the carrier spectra.

\section{Scattering mechanisms}

For the GBL-RIL hererostructures under consideration, we have  the following formulas:
 
\begin{equation}\label{eq5}
|V(\hbar{\bf q})|^2 = \biggl[\frac{2\pi\,e^2}{\kappa(q + q_{TF})}\biggr]^2
[\Sigma_A + \Sigma_{RCL}\exp(-2qd)],
\end{equation}
for the discrete acceptors in the GBL and the discrete acceptors and donors in the RIL
and
\begin{multline}\label{eq6}
|V(\hbar{\bf q})|^2 = \biggl[\frac{2\pi\,e^2}{\kappa(q + q_{TF})}\biggr]^2\\
 \times [\Sigma_A + Z_c\Sigma_{RCL}\exp(-2qd - q^2l^2/2)],
\end{multline}
for the point charged defects  (acceptors) in the GBL and the correlated
clusterized acceptors and donors in the RIL, respectively.
Here $q = |{\bf p} - {\bf p}^{\prime}|$, $\Sigma_{RCL} = \Sigma_{A,R} + \Sigma_{D, R}$ 
is the net density
of the charge impurities of both types located in the RIL
(therefore  the cluster density in the "clusterized" RIL is equal to  $\Sigma_{RCL}/Z_c$ with
$\Sigma_{RCL} >\Sigma_A$),
$\kappa$ is  the effective dielectric constant (which is the half-sum
of the dielectric constants of the materials surrounding the GBL,  i.e., of the spacer and the substrate), $q_{TF} = (4e^2m/\hbar^2\kappa)$~is the Thomas-Fermi screening wave number~\cite{38,39,40}, which
is independent of the carrier density.
Deriving Eq.~(6), we have taken into account that the density of the charged clusters in the clusterized RCL
 is $Z_c$-times smaller than the donor and acceptor densities,
but  the  cross-section of the carrier scattering on the clusters  is proportional to $Z_c^2e^4$.  
The factor $\exp(-2qd)$ in Eqs.~(5) and (6) is associated with the remote position of a portion of the charged scatterers (see, for example~\cite{31}). Equation~(7) corresponds to the Gaussian spatial  distributions of the clusters that leads to the appearance of the factor $\exp(-q^2l_c^2/2)$
(compare with Refs.~\cite{29,41,42,43}).

One can see that the scattering matrix element explicitly  depends on the 
 background dielectric constant $\kappa$ and via the dependence of $q_{TF}$ on $\kappa$.

\begin{figure}[t]
\centering
\includegraphics[width=7.0cm]{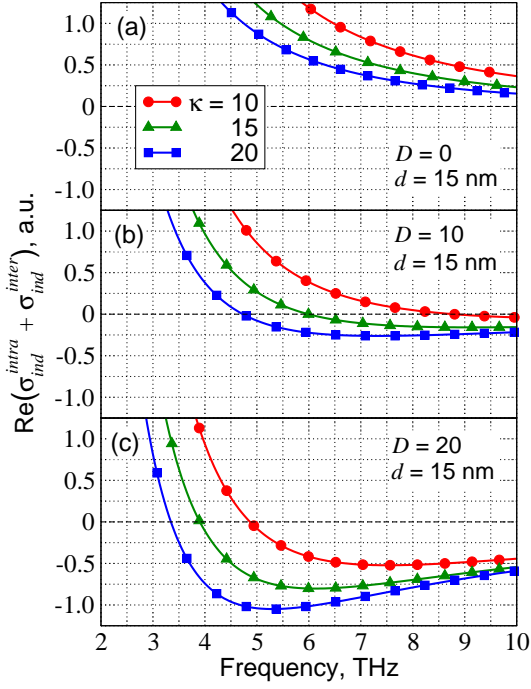}
\caption{ 
Frequency dependences of ${\rm Re}~(\sigma_{ind}^{intra} + \sigma_{ind}^{inter})$
calculated for GBL-RIL heterostructures with spacer thickness  $d = 15$~nm, different dielectric constants ($\kappa = 10, 15,$ and 20), and different doping parameters $D$:
(a) $D = 0$, (b) $D = 10$, and (c) $D = 20$.} 
\end{figure}
\begin{figure}[th]
\centering
\includegraphics[width=7.0cm]{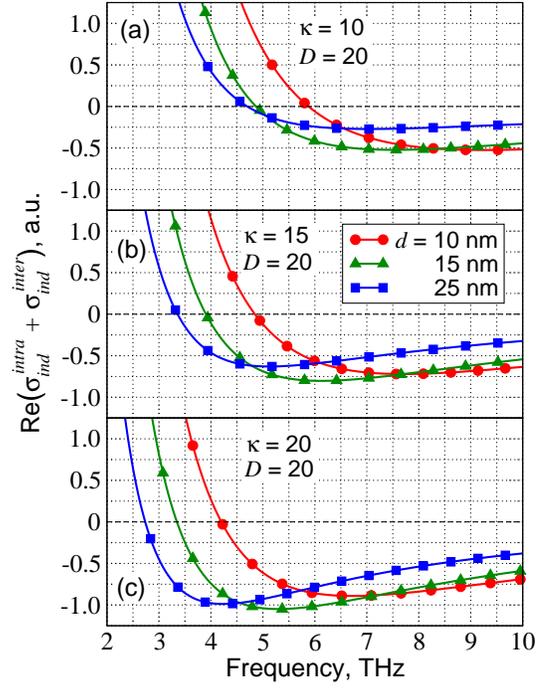}
\caption{ The same as in Fig.~3 but for doping parameter $D = 20$, different spacer
thicknesses ($d = 10, 15,$ and 25~nm), and different dielectric constants: (a) $\kappa = 10$, (b) $\kappa = 15$, and (c) $\kappa = 20$.
}
\end{figure}

\section{Net THz conductivity}

In the following we consider  the most practical case of
the temperature,   frequency range, and pumping in which $\hbar/\tau < \hbar\omega,  T < \varepsilon_F$.
For $T = 300$~K,  the frequency range $\omega/2\pi = 1 - 10$~THz, and $\varepsilon_F \gtrsim 50$~meV, these inequalities 
 are satisfied
when $\tau \gtrsim 0.05 - 0.5$~ps.

For GBL structures without the RIL ($\Sigma_{RIL} = 0$) and the totally screened charges in the GBL,
in which ${\rm Re}~\sigma_{ind}^{intra} = \overline {\sigma_{ind}^{intra}}$ and ${\rm Re}~\sigma_{ind}^{inter} = \overline {\sigma_{ind}^{inter}}$, invoking Eqs.~(3) - (6), one obtains

 \begin{equation}\label{eq7}
{\rm Re}~\overline{\sigma_{ind}^{intra}} \simeq 2\frac{e^2\Sigma}{m\tau}\frac{1}{\omega^2} 
= 2\sigma_Q \biggl(\frac{\omega_D}{\omega}\biggr)^2,
\end{equation}
where $\omega_D = \sqrt{4\hbar\Sigma/m\tau} = \sqrt{8\varepsilon_F/\pi\hbar\tau}$ is the Drude frequency. This quantity characterizes the relative strength of the Drude processes (absorption). 

As the above takes place, 
Eqs.~(3) and (4) lead to 

 \begin{equation}\label{eq8}
\frac{{\rm Re}~\overline{\sigma_{ind}^{inter}}}{{\rm Re}~\overline{\sigma_{ind}^{intra}}} \simeq -\frac{\hbar\omega}{4\varepsilon_F}.
\end{equation}

To avoid cumbersome   calculations and obtain
transparent formulas suitable for a qualitative analysis, we use the mean value theorem for the  integrals
over $d{\bf q}$. As a result, we arrive at
the following simplified formulas for ${\rm Re}~\sigma_{ind}^{intra}  + {\rm Re}\sigma_{ind}^{inter})$:

\begin{multline}\label{eq9}
{\rm Re}~(\sigma_{ind}^{intra}  + \sigma_{ind}^{inter})\\
\simeq   {\rm Re}~\overline{\sigma_{ind}^{intra}}
\biggl\{\displaystyle \int_{Q_{min}}^{Q_{max}} 
dqq\frac{[1 +  D\exp(-2 qd - q^2l^2/2)]}{(Q_{max}^2 - Q_{min}^2)(q/q_{TF} +1)^2}
\\
 -\frac{\hbar\omega}{4\varepsilon_F}\displaystyle \int_{q_{min}}^{q_{max}} 
 dqq\frac{[1 + D\exp(-2 qd - q^2l^2/2)]}{(q_{max}^2 - q_{min}^2)(q/q_{TF} + 1)^2}
\biggr\}.
\end{multline}
Here 
  $D = Z_c(\Sigma_{RCL}/\Sigma_A) = Z_c(\Sigma_{D,R} + \Sigma_{A,R})/
(\Sigma_{D,R} - \Sigma_{A,R}) \simeq Z_c(\Sigma_{D,R} + \Sigma_{A,R})/
\Sigma_A \geq Z_c \geq 1$ is the doping parameter ($Z_c = 1$ and $l = 0$ for the RGL with the point charges
(acceptors and donors) and $Z_c > 1$ and $l =l_c$  in  the case of RIL with 
the clusterized acceptors and donors),
$\hbar Q_{max} \simeq 2\sqrt{2m\varepsilon_F}(1 - \hbar\omega/4\varepsilon_F)$, 
$\hbar Q_{min} \simeq 2\sqrt{2m\varepsilon_F}(\hbar\omega/4\varepsilon_F)$,
$\hbar q_{max} \simeq 2\sqrt{m\hbar\omega}$,$q_{min} \simeq 0$. 
The latter quantities follow from the conservation laws for the indirect intraband and interband radiative transitions.
%
%
In particular, Eq.~(9) gives rise to

\begin{multline}\label{eq10}
\frac{{\rm Re}~(\sigma_{ind}^{intra} + \sigma_{ind}^{inter})}{{\rm Re}\sigma_{ind}^{intra}}
\simeq  1\\
 -\frac{1}{2}\,
\frac{\displaystyle \int_{q_{min}}^{q_{max}} 
 dqq\frac{[1 + D\exp(-2 qd - q^2l^2/2)]}{(q/q_{TF} + 1)^2}}
{\displaystyle \int_{Q_{min}}^{Q_{max}} 
dqq\frac{[1 +  D\exp(-2 qd - q^2l^2/2)]}{(q/q_{TF} + 1)^2}}.
\end{multline}

For sufficiently pure GBLs and relatively strong RIL doping  (with nearly complete  compensation of the donor and acceptors), the doping parameter parameters $D$ can be fairly large. In this case, the contribution of the carrier scattering on
the remote impurities to the net scattering rate can be dominant. In such a case, the absolute value of the net THz conductivity
$|{\rm Re}\sigma|$ can markedly exceed the contribution solely from  the direct interband transitions, i.e.,
$|{\rm Re}\sigma| > 2\sigma_Q$.

Considering Eqs.~(1), (2), and (9), we arrive at the following formula for the net THz conductivity:

\begin{multline}\label{eq11}
\frac{{\rm Re}~\sigma}{2\sigma_Q} \simeq  \displaystyle\tanh\biggl(\frac{\hbar\omega - 2\varepsilon_F}{4T} \biggr)\\
+  \biggl(\frac{\omega_D}{\omega}\biggr)^2
\biggl\{\displaystyle \int_{Q_{min}}^{Q_{max}} 
dqq\frac{[1 +  D\exp(-2 qd - q^2l^2/2)]}{(Q_{max}^2 - Q_{min}^2)(q/q_{TF} +1)^2}
\\
 -\frac{\hbar\omega}{4\varepsilon_F}\displaystyle \int_{q_{min}}^{q_{max}} 
 dqq\frac{[1 + D\exp(-2 qd - q^2l^2/2)]}{(q_{max}^2 - q_{min}^2)(q/q_{TF} + 1)^2}
\biggr\}.
\end{multline}

\section{Results and Analysis}

\begin{figure}[t]
\centering
\includegraphics[width=7.0cm]{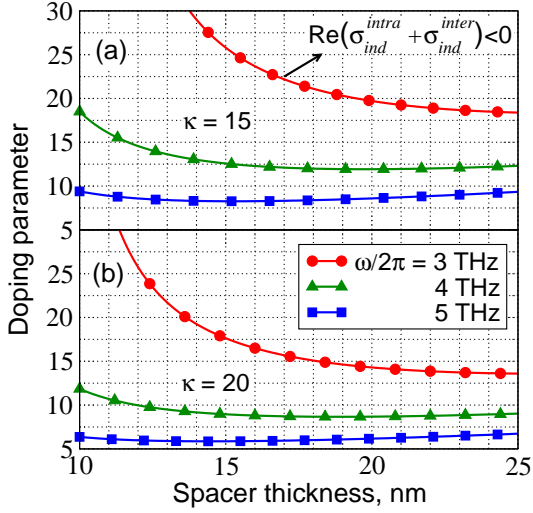}
\caption{Areas in the $D - d$ plane where  ${\rm Re}~(\sigma_{ind}^{intra} + \sigma_{ind}^{inter}) > 0$
(areas below lines corresponding to given frequencies) and ${\rm Re}~(\sigma_{ind}^{intra} + \sigma_{ind}^{inter}) < 0$ (above these lines) at  different values of dielectric constant  $\kappa$: (a) $\kappa = 15$ and (b) $\kappa = 20$.
}
\end{figure}

\begin{figure}[t]
\centering
\includegraphics[width=7.0cm]{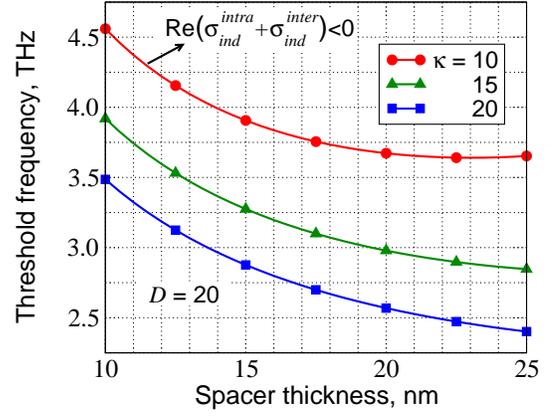}
\caption{ 
Threshold frequency $\omega_0/2\pi$  corresponding to ${\rm Re}~(\sigma_{ind}^{intra} + \sigma_{ind}^{inter}) = 0$
as a function of the spacer thickness $d$ calculated for $D = 20$ and different values of  dielectric constant $\kappa$. AS in Fig.~5, areas above the pertinent line are related to ${\rm Re}~(\sigma_{ind}^{intra} + \sigma_{ind}^{inter}) < 0$.
}
\end{figure}

Using Eqs.~(9) and (10), one can  analyze the conditions when the interband radiative processes surpass the intraband one's,
i.e., when  ${\rm Re}~(\sigma_{ind}^{intra}  + \sigma_{ind}^{inter}) < 0$.

For the practical  GBL-RIL heterostructures with 
the  HfO$_2$ spacer and the substrate material like  Si and  SiC,
 the effective dielectric constant (which is the half-sum
of the dielectric constants of the spacer and the substrate) is  in the range $\kappa = 10 - 20$.
In the calculations we assume set $m = 4\times 10^{-29}$~g and $\varepsilon_F = 60$~meV ($\Sigma \sim 2.5\times 10^{12}$~cm$^{-2}$).

Figures~3 and 4 show the dependences of ${\rm Re}~(\sigma_{ind}^{intra} + \sigma_{ind}^{inter})$ on the frequency $f = \omega/2\pi$ calculated using Eq.~(9) with $l = 0$ (uncorrelated impurities in the RIL) for different values of 
the  doping parameter $D$,  the spacer thickness  $d$, and effective dielectric constant $\kappa$. 
As seen from Fig.~3(a), in the absence of the RCL ($D = 0$), the indirect intraband transitions
prevail over the indirect interband transitions. In such a case,
the indirect transitions lead to the photon absorption  in the entire frequency range,
although the photon emission due to the interband transitions partially compensates the Drude absorption.
This compensation is enhanced in the heterostructures with the doped RIL, i.e., for $D > 0$
[see Figs. 3(b) and 3(c)]: an increase in the doping parameter $D$ leads to the appearance of the frequency range where ${\rm Re}~(\sigma_{ind}^{intra} + \sigma_{ind}^{inter}) < 0$.  The latter range becomes wider (it starts from lower frequency) with increasing dielectric constant $\kappa$. These features of the transformation of the ${\rm Re}~(\sigma_{ind}^{intra} + \sigma_{ind}^{inter})$ frequency dependence are attributed to  reinforcing of the long-range carrier
scattering (with increasing $D$) and smoothing of the scattering potential (due to a stronger screening at higher $\kappa$). Both these effects result in an increase of the relative role of the indirect interband transitions. The scattering potential smoothing associated with an increase in the spacer thickness $d$ also beneficial for the indirect interband transitions is seen in Fig.~4 as well.

Figure~5 shows  the $D - d$ plane where  ${\rm Re}~(\sigma_{ind}^{intra} + \sigma_{ind}^{inter}) > 0$
(below the $D$ versus $d$ lines)
and ${\rm Re}~(\sigma_{ind}^{intra} + \sigma_{ind}^{inter}) < 0$ 
(above these lines) calculated considering  Eq.~(10 )at  different values of  $\kappa$ and frequency
$f = \omega/2\pi$. As seen,  the ranges of parameters $D$ and $d$ where the quantity ${\rm Re}~(\sigma_{ind}^{intra} + \sigma_{ind}^{inter})$ is negative
are markedly wider
in the case of relatively large dielectric constant $\kappa$ and the frequency $f$.

\begin{figure}[t]
\centering
\includegraphics[width=7.5cm]{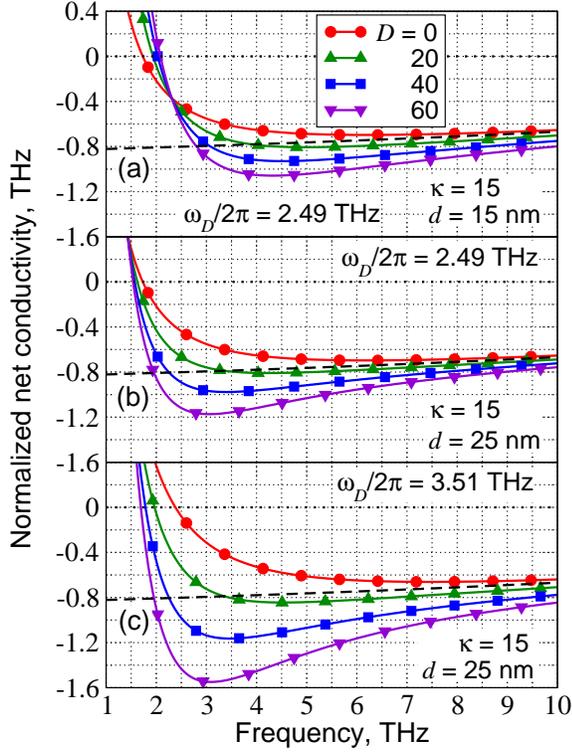}
\caption{ 
Frequency dependence of the normalized net THz conductivity ${\rm Re}\sigma/2\sigma_Q$
of GBL-RIL heterostructures with $\kappa = 15$ and different values of doping parameter for:
(a) $d = 15$~nm and  $\omega_D/2\pi = 2.49$~THz ($\tau = 1$~ps), (b)
$d = 25$~nm and $\omega_D/2\pi = 2.49$~THz($\tau = 1$~ps) , 
and (c) $d = 25$ and $\omega_D/2\pi = 3.51$~THz ($\tau = 0.5$~ps). Lines marked by circles 
 correspond to  normalize net THz conductivity  without RIL ($D = 0$).
Dashed lines show  normalize THz conductivity due to solely
 direct interband transitions ${\rm Re\sigma_d/2\sigma_Q}$.
}
\end{figure}

\begin{figure}[t]
\centering
\includegraphics[width=7.0cm]{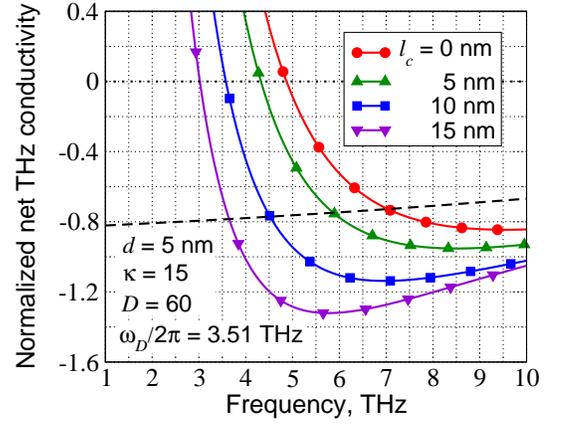}
\caption{ 
Frequency dependence of the normalized net THz conductivity ${\rm Re}~\sigma/2\sigma_Q$
of GBL-RIL heterostructures with clusterized impurities,  $\kappa = 15$, $\omega/2\pi = 3.51$~THz, $d = 15$~nm, $D = 60$, and different values of the cluster size $l_c$.
}
\end{figure}

Figure~6 shows the threshold frequency $\omega_0/2\pi$  corresponding to ${\rm Re}~(\sigma_{ind}^{intra} + \sigma_{ind}^{inter}) = 0$
as a function of the spacer thickness $d$ calculated for $D = 20$ and different values of  $\kappa$.

Figure~7 shows the frequency dependences of the normalized net THz conductivity 
${\rm Re}~ \sigma/2\sigma_Q$ calculated using Eq.~(11) for $T = 300$~K and different values of the 
doping parameter $D$, dielectric constant  $\kappa = 15$,  and Drude frequency $\Omega_D/2\pi$. The
Drude frequencies are   chosen to be $\omega_D/2\pi = 2.49$ and $3.51$~THz. At $\varepsilon_F = 60$~meV, this corresponds to the carrier momentum relaxation time $\tau = 1$~ps and 0.5~ps, respectively.
The latter values of $\tau$ corresponds to the carrier mobility in the GBL  without the RIL, i.e., with $D = 0$,
equal to  $\mu = 40,000$~cm$^2$/V s and 20,000~cm$^2$/V s. The large values of  $D$ imply rather strong doping of the RIL. For example, if $\Sigma_A = 1\times10^{11}$~cm$^{-2}$ (that is consistent with the above value
of $\tau$), the parameter $D$ is equal to 60 if
$\Sigma_{D,R} = 3.05\times 10^{12}$~cm$^{-2}$ and $\Sigma_{D,R} = 2.95\times 10^{12}$~cm$^{-2}$.
g.
As seen in Fig.~7,  the net THz conductivity is negative
in a certain range of frequencies (above 1.5 THz) even for the case $D = 0$, i.e., even in the GBL structures without the RIL. This is because although  the indirect intraband transitions (Drude transitions) prevail over
the indirect  interband transitions, the negative contribution of the direct interband transitions
enable the negativity of the net conductivity.  The latter is because
the contribution of the direct transitions decreases with increasing frequency
much slowly (see the dashed line in Fig.7) than the contribution  due to the Drude processes.
However, in the GBL-RIL heterostructures with sufficiently strong doping (large $D$) the absolute value
of the net THz conductivity is pronouncedly larger than in the similar heterostructures without the RIL.
Moreover, the substantial contribution of the indirect interband transition can result in
$|{\rm Re}~\sigma|$ markedly exceeding its values $|{\rm Re}~ \sigma_d|$ associated with solely direct interband transitions 
(see the curves s for $D = 40,$ and 60 and the dashed lines in Fig.~7) and even exceed
the maximum value of the latter $2\sigma_Q$ (see the curve for $D = 60$). 
The comparison of the frequency dependences in Fig.~7 for $D = 0$ and $D = 60$ reveals that 
the RIL can provide a substantial increase in the absolute value 
 of the THz conductivity [up to 4 - 6 times increase  in the range $\omega/2\pi = 3 - 5$~THz
,compare the curves for $D = 60$ and $D = 0$ in Fig.~7(c)] and, hence,
the corresponding  enhancement of the laser THz gain (for the parameters under consideration).
Such an increase can be even more large at somewhat lower frequencies.

The comparison of Figs.~7(b) and 7(c) also demonstrates that at larger Drude frequencies, i.e., at  shorter carrier momentum relaxation time $\tau$ (and the same parameter $D$), the sag of
 the net THz conductivity frequency dependences becomes deeper. This is attributed to an increase
 of the intensity of the indirect processes (both intraband and interband)
when $\tau$ decreases.

The plots in Fig.~8 were obtained using Eq.~(11) with $l = l_c \neq 0$.
Figure~8  shows   the frequency dependences of the normalized net THz conductivity
in the GBL-RIL heterostructures with  clusterized (correlated)  impurities in the RIL.
As seen from  Fig.~8 , for relatively small spacer thickness ($d = 5$~nm), the clusterization leads to an additional enhancement of the indirect interband transitions, reinforcing the  negative THz conductivity, particularly with an increasing
cluster size. However, at relatively large spacer thicknesses ($d \gtrsim 10$~nm), an increase in the absolute value of the net THz conductivity with increasing size of the clusters is insignificant,
since in this case, $D$ is determined not only by the doping of the RIL but the cluster charge $Z_c$.
This means that the chosen value $D = 60$ can correspond to different densities $\Sigma_{D,R}$
and $\Sigma_{A,R}$ depending on $Z_c$. The latter, in turn, depends of the degree of the compensation
of donors by acceptors in the clusters.

\section{Discussion}

If the RIL doping does not compensate the acceptor system in the GBL, the electron and hole
quasi-Fermi energies $\varepsilon_{F,e}$ and $\varepsilon_{F,h}$ are generally not equal to each other
even in the pumping conditions. However, at sufficiently strong pumping when the density of the injected (or optically pumped) carriers   $\Sigma > \Delta \Sigma = \Sigma_{D,A} - \Sigma_{D,A} - \Sigma_{D,A}$,
Eqs.~(2), (9), and (10) are still approximately valid until $\hbar\omega/2  < {\rm min} \{\varepsilon_{F,e},  \varepsilon_{F,h}\}$.

The above model disregarded the opening of the band gap in the GBL under the transverse electric field arising due to the impurity charges  in the RIL and GBL 
Despite a fairly complex pattern of the opening gap in GBLs by the electric field and doping~\cite{41,42},
the pertinent energy gap $\Delta_g$ can roughly be estimated in the manner as it was done in~\cite{35}.
In the absence of the electron and hole injection from the side contacts,
the carrier density in the GBL is rather small due to the compensation of the acceptor charges in the GBL
and the donor and acceptor charges in the RIL (see, Sec.II). In this case, 
$\Delta_g \sim eE_{\perp}d_{GBL}$, where $E_{\perp} = 4\pi\,e \Sigma_A/\kappa$ is the transversal electric field created by the charged impurities and $d_{GBL} \simeq 0.36$ nm is 
the spacing between GLs in the GBL, so that $\Delta_g \sim 4\pi\,e^2\Sigma_Ad_{GBL}/\kappa$.
Under the pumping conditions, the 2D electron and hole gases partially screen the electric field in the  GBL structure. As a result, one obtains $\Delta_g \sim eE_{\perp}d_{GBL}^{eff}$ with $d_{GBL}^{eff} < 
d_{GBL}$~\cite{35}. Taking into account the screening of the transversal  electric field $E_{\perp}$ by both electron and hole components, we
obtain the following estimate: $\Delta_g \sim 4\pi\,e^2\Sigma_Ad_{GBL}/\kappa[1 + (8d_{GBL}/a_B)]$,
where $a_B = \kappa \hbar^2/me^2$ is the Bohr radius. Assuming $a_B = 10 - 20$~nm and $\Sigma_A = (5 - 10)\times 10^{11}$~cm$^{-2}$, we obtain $\Delta_g \sim 1.5 - 5 $~meV. These values 
are in line with those 
estimated in~\cite{35} and extracted from the experimental data~\cite{44,45}.
This estimate validates our model 

In the case of the clusterized charged impurities, the long range variations of the potential associated with
the clusters lead to the variation of the band gap. The latter variations reinforce the electron and hole scattering  on the clusters, but this effect is not really essential.

In Eqs.~(3) and (4), we have disregarded the indirect radiative processes associated with electron-hole scattering in the GBL. The point is that the probability of such processes is proportional
to $[2\pi\,e^2/\kappa(q + q_{TF})]^2\Sigma\eta$, where (in the degenerated two-dimensional electron and hole gases),
the factors $\eta < 1$ is the  fraction of electrons and holes effectively participated
in the scattering processes   $\eta = T/\varepsilon_{F}$ at $\hbar\omega < T$,  so that $\Sigma\eta \simeq \Sigma_T = 2mT/\pi\hbar^2$. 
The values of  $q_{TF}$ and  $\Sigma\eta $ are independent of
the electron and hole densities and, hence, of the pumping conditions. This is  because of the virtually constantdensity of state.
At rooom temperatures, the  value of $\Sigma_T$ can be of the order of or less than $\Sigma_A$
and much smaller than $\Sigma_{RIL}$. In the former case, one needs to replace the quantity $D$ in equations~(9) and (10) by $D^* = D/(1 + 2\Sigma_T/\Sigma_A)$. 

\section{Conclusions}
We proposed to use  the GBL-RIL heterostructure with the population inversion due to
the electron and hole injection 
as the active region of interband THz lasers and demonstrated that the incorporation of the RIL
enables a substantial reinforcement of the effect of negative THz conductivity and the laser THz gain.
This is associated with  the domination of the indirect interband transitions in the GBL over
the indirect intraband transitions (resulting in the Drude absorption) when the carrier scattering
on a long-range potential prevails. As shown, the latter can be  realized due to the remote doping
and enhanced by the clusterization of impurities. 

Using a simplified device model, we calculated the THz conductivity of the GBL-RIL heterostructures
as function of the frequency for different structural parameters (dopant densities, spacer thickness,
 impurity cluster size, and the injected carrier momentum relaxation time). 
We found that the absolute value of the THz conductivity  in the GBL-RIL heterostructures with a sufficiently highly doped RIL separated from the GBL by a spacer layer
of properly chosen thickness 
can exceed in several times the pertinent value in the GBL-heterostructures without the RIL. 
Thus, the remotely doped  GBL heterostructures
 can be of interest for applications  as the active media in the  THz lasers.

\section*{Acknowledgments}

The authors are grateful to D. Svintsov and V. Vyurkov for  useful discussions and information.
The work was supported by 
the Japan Society for Promotion of Science (Grant-in-Aid for Specially Promoted Research $\#$ 23000008) and by 
the Russian Scientific Foundation (Project $\#$14-29-00277).
The works at UB and RPI were  supported  by the US Air Force
award $\#$ FA9550-10-1-391  and by the US Army Research Laboratory
Cooperative Research Agreement, respectively.

\end{document}